# Critical Empirical Study on Black-box Explanations in AI

*Short Paper*

**Jean-Marie John-Mathews**
Université Paris-Saclay, Univ Evry, IMT-BS, LITEM
91025, Evry, France
jean-marie.john-mathews@imt-bs.eu

## Abstract

*This paper provides empirical concerns about post-hoc explanations of black-box ML models, one of the major trends in AI explainability (XAI), by showing its lack of interpretability and societal consequences. Using a representative consumer panel to test our assumptions, we report three main findings. First, we show that post-hoc explanations of black-box model tend to give partial and biased information on the underlying mechanism of the algorithm and can be subject to manipulation or information withholding by diverting users' attention. Secondly, we show the importance of tested behavioral indicators, in addition to self-reported perceived indicators, to provide a more comprehensive view of the dimensions of interpretability. This paper contributes to shedding new light on the actual theoretical debate between intrinsically transparent AI models and post-hoc explanations of black-box complex models – a debate which is likely to play a highly influential role in the future development and operationalization of AI systems.*

## Introduction

Artificial Intelligence (AI) systems are increasingly used in a wide variety of sectors such as human resources, banking (Gan et al. 2020), and legal services, for tasks such as job candidate screening (Liem et al. 2018), medical diagnosis (Kononenko 2001), etc. As these systems significantly impact businesses and organizations, they spawn many new challenges. Algorithm biases, discrimination or privacy issues have been identified in various AI applications, such as predictive models in justice (Larson et al. 2016), search engines (Kay et al. 2015), and predictive models in healthcare (Obermeyer et al. 2019). Consequently, ethics of AI have become a big challenge for society, and the responsible AI movement has spread to most countries and universities, policy makers, companies and non-profit organizations.

While building interpretable AI systems can be seen as a solution for accountable and trustworthy algorithms, it is often claimed that some AI algorithms are inherently black boxes (Jang 2019). Some models, such as Deep Learning, are indeed high-dimensional, use non-linear transformations, and are not based on formal probabilistic theories that ensure their mathematical validity (Evermann et al. 2017; Mahmoudi et al. 2018). As a consequence, a large and very active community of computer scientists, called XAI (Explainable Artificial Intelligence), have recently developed many analytics tools to a posteriori create local explanations from black-box models. These tools are often called post-hoc explanations. One of the popular methods in post-hoc explanations is feature relevance explanation techniques, such as Shapley explanations (Lundberg et al. 2017) (see Appendix C), which aim to describe the functioning of a black-box model by measuring the influence that each feature has in the prediction to be explained (Barredo Arrieta et al. 2020). Another popular post-hoc explanation techniques family is heuristics for creating counterfactual examples by modifying the input of the model (Martens 2014). Many post-hoc explanations tools are being created by IT companies like IBM or Sheldon ("AI Explainability 360"; "SeldonIO/Alibi) to provide post-hoc explanations of potentially black-box models.

Other authors criticize the use of post-hoc explanations. They call for the development of intrinsically transparent models (i.e. non black-box model) since they show they can often perform just as well as black-box models (Rudin 2019). Some even claim that post-hoc explanations act as a pretext for companies to continue to produce non-accountable algorithms and avoid strong regulation (Martin 2018)

Is it always necessary to build inherently intelligible models? Or is it preferable to create models that are possibly more efficient while providing an a posteriori explanation by an additional explanation method? Is it possible to empirically compare the two alternatives on some specific use cases?

This recent technical controversy between intrinsically transparent ML model and post-hoc explanations of black-box models (black-box explanations) will play a fundamental role in the development of a trustworthy AI. While Machine Learning (ML) algorithms are more and more complex and post-hoc explanations of black-box models are more and more put forward by companies, it is highly important to carry out an empirical evaluation to compare black-box explanations and transparent models This paper provides practical limitations of black-box explanations in terms of interpretability. While the debate between transparent models and post-hoc explanations of black-box models is often discussed theoretically, we empirically test this controversy. To do so, we gathered a representative panel of the population to test these different scenarios in a real-world setting. Without focusing only on self-reported subjective indicators of interpretability, we compared perceptive and behavioral indicators to provide a more comprehensive view of the dimensions of interpretability. We think that the opposition between self-reported indicators and tested behavioral indicators is key to understanding the pain points of post-hoc explanations of black-box models. This leads us to make recommendations to design ML algorithms that do not undermine the agency of non-expert citizens.

## Theoretical background

Interpretability is a difficult concept to define (Lipton 2018). As in previous works (Doshi-Velez and Kim 2017; Poursabzi-Sangdeh et al. 2018), we consider that interpretability is a latent property composed of several dimensions. We take into account the three following dimensions of explainable AI in the literature: understandability, actionability and, generalizability. These dimensions are chosen for their frequency of appearance in the literature review of XAI metrics (Hoffman et al. 2018) but also for their ability to be tested in practice (see next section).

Understandability denotes the characteristic of a model to make a human understand its mechanism without explaining its internal structure (Barredo Arrieta et al. 2020). For high-dimensional non-linear machine learning models (e.g. Deep Learning techniques), post-hoc explanations are specifically adapted for understandability since they aim to explain the algorithmic decision without revealing its underlying mechanism. One of the interpretability techniques that look for understandability is feature relevance explanation methods with the Shapley (Lundberg et al. 2017) method (see Appendix C for an example). These methods compute the sensitivity for each feature upon the output of the model. As a result, they give the features that are the most important when producing the output of the model.

Actionability is another important dimension of interpretability (Krause et al. 2016). Explanations that are actionable give the user guidelines for the desired outcome. Whilst some features represent properties that cannot be influenced (e.g. an individual's age), others capture characteristics that can be adjusted (e.g. credit duration). Thanks to these actionable explanations, adjustable features can be modified so that a prediction gives the desired outcome when the mutated instance is put back to the model.

Generalizability or completeness is also another dimension of interpretability. It denotes the characteristic of a model to make a human generalize well beyond the particular case in which the algorithmic decision was initially produced. Generalizability is important to apply an insight learnt from one explanation to other similar cases (Miller 2019).

## Data and Methods

The objective of this study is to analyze how consumers react to different types of explanations following a hypothetical refusal of credit from an AI algorithm in a real-world setting.

We designed a randomized experiment on 400 people with 4 scenarios, using a consumer panel collected in France. Each scenario was tested on a random sample of 100 adults that is representative of the French population, using statistical quotas on gender and age. We also verified that none of the population, from each of the 4 scenarios, is statistically distributed differently in terms of education level and socio-professional category, compared to the total population. For each scenario, we asked the participants to imagine that they applied to a bank for a credit and that the bank is using an artificial intelligence algorithm to accept or refuse the application. For this, we specified that the automatic algorithm uses 6 pieces of information, namely credit amount (1500 euros), credit duration (26 months), installment rate in percentage of disposable income (4%), number of years in employment (5 years), property (car ownership) and the number of credit application approvals in the past at the same bank (only one past approval).

These 6 features are themselves extracted from a representative case in the German Credit Scoring database from the UC Irvine Machine Learning repository using the same design as in (Lu et al. 2019). This dataset consists of 1000 individual profiles, each with 20 features and categorized as either "good" or "bad" credit risks. From the 1000 individuals in the German Credit Scoring database [37], we built two Machine Learning algorithms to decide whether to accept or refuse the credit application: a logistic regression with integer coefficients and a 2-layer feedforward neural network of 10 neurons in each layer. Using these two Machine Learning algorithms, we constructed 4 different scenarios corresponding to each interpretability method. For each scenario, we start by introducing the case, then give the input variables used as input by the AI algorithm, and inform the participant that his or her credit has been refused by the algorithm (see Appendix A). Comparing post-hoc explanations of black-box models with transparent algorithm explanation is a difficult task since both are using different models. To control this difference, we chose a case from the German credit scoring that gives the same post-hoc explanation for both black-box model and transparent model[1].

**Scenario 1 (No explanation – Baseline):** In the first scenario, we do not provide any further explanation other than the aforementioned introduction. This scenario will be used as a baseline to measure the effects of the other scenarios.

**Scenario 2 (Transparent algorithm):** In this second scenario, we provide a points-based system using the integer coefficient of the logistic regression model to explain why the decision was negative. To do so, we assign to each feature a certain number of points obtained from the coefficients of the logistic regression (Rudin 2019). If the sum of all the points exceeds a certain threshold, then the bank loan is accepted (see Appendix B). The linearity of the logistic regression makes it easy to represent the decision rule as a sum of points. That is why, in the XAI literature, logistic regression with few variables is considered as a transparent or intrinsically interpretable algorithm (Rudin 2019). We will refer to Scenario 2 as the transparent algorithm scenario.

**Scenario 3 (Post-hoc Shapley):** In the third scenario, we rather use the neural network to take the algorithmic decision. Contrary to the logistic regression, since the neural network cannot be represented as a simple linear additive model, we choose to provide a post-hoc explanation of the decision using the Shapley method (a feature relevance method around local input). In other words, we inform participants of the importance of each feature with respect to the negative decision made by the black-box algorithm (see Appendix C). We will refer to Scenario 3 as the Shapley post-hoc explanation scenario.

**Scenario 4 (Post-hoc counterfactual):** In this scenario, we use the same black-box neural network to make the decision to refuse credit. However, instead of giving feature relevance for each variable, we give counterfactual explanations ('What if' explanation) for the two actionable features, i.e. credit duration and credit amount. To generate counterfactual explanations, we find, for both variables, the threshold that turn the negative decision to a positive decision. Applied to the two actionable features, this method gives two different counterfactual explanations that can help participant to know how to modify his or her demand to change the algorithm output (see Appendix D).

---

[1] We compare explanations techniques as a whole, not just the ways of presenting the same explanation. Please note that, apart from having the same behavior locally around the selected case, deep learning and logistic regression have very similar performance metrics more globally on test sets of German Credit Scoring (AUC=0.8). This is in line with Rudin (2019): complex models do not necessarily have better performance.

To measure the perception of these different scenarios, we ask the participants to rate their agreement with several statements (second column in Table 1) about the algorithm, using a scale from "1" (strongly disagree) to "7" (strongly agree). These statements build 5 perceived indicators: understandability, complexity, actionability, generalizability and trust. Since these perceived indicators are subjective, we choose to complement them using behavioral performance scores in test assessments for the dimensions of understandability, actionability and generalizability (third column in Table 1). Since our scenarios are based on algorithms built on real world data, each assessment test has a right answer, and it becomes possible to calculate the average number of right answers for each of these three tested dimensions. Thus, in addition to the perceived dimensions of understandability, actionability and generalizability, we also have three additional indicators that we call the tested or behavioral dimensions of interpretability. To avoid biases on perceived dimensions, we first ask perception questions before behavioral tests. We begin the interview by asking the participants about their level of knowledge in artificial intelligence, the banking sector, and their level of education (graduation from higher education). Finally, to make sure that the participants answered the questions carefully, we asked them attention questions about their reading of the statement, which allowed us to filter out 10% of the participants that did not read the statements carefully.

| | Perceived dimension | Tested dimension |
|---|---|---|
| **UNDERSTANDABILITY** | You understand the loan algorithm | *Did your installment rate work in favor or against your credit application?*<br>1- My 4% installment rate worked in favor of my request<br>2- My 4% installment rate worked against my request **(right answer)**<br>3- I don't know<br><br>*Did your number of years of employment work in favor or against your credit application?*<br>1- My 5-years of employment worked in favor of my request **(right answer)**<br>2- My 5-years of employment worked against my request.<br>3- I don't know |
| **ACTIONABILITY** | If you have the possibility to reapply, you have the means to modify your credit demand to obtain the bank loan | *Suppose that you have the possibility to reapply. How would you modify the credit amount to increase your chances of approval?*<br>1. Modify the credit amount to 3500 euros **(right answer)**<br>2. Modify the credit amount to 1000 euros<br>3. I don't know<br><br>*Suppose that you have the possibility to reapply. How would you modify the credit duration to increase your chances of approval?*<br>1. Modify the credit duration to 40 months<br>2. Modify the credit amount to 8 months **(right answer)**<br>3. I don't know |
| **GENERALIZABILITY** | You can predict the output of the loan algorithm for another loan demand | *What would the result of the algorithm be if you had a life insurance as property?*<br>1- My loan application would have an even lower chance of being accepted **(right answer)**<br>2- My loan application would have a better chance of being accepted.<br>3- I don't know<br><br>*What would the result of the algorithm be if you had obtained more bank credit acceptances in the past?*<br>1- My loan application would have an even lower chance of being accepted.<br>2- My loan application would have a better chance of being accepted **(right answer)**<br>3- I don't know |

| Table 1. | Questions put to participants to measures perceived and tested dimensions |
|---|---|

## Preliminary Results

Even if all forms of explanation increase understanding with respect to the situation without explanation, counterfactual explanations do not significantly increase understanding contrary to transparent and Shapley explanations (Table 2). Transparent explanation is perceived to be more understandable than Shapley explanation, whereas Shapley explanation scores higher on understandability assessment tests.

Table 2 shows that perceived actionability is significantly improved by the presence of an explanation. However, when performing the actionability tests, we see that actionability is almost the same compared to the situation without explanation. Therefore, it seems that Shapley explanations are wrongly perceived as being actionable. We can give an explanation for this fact. Feature relevancy explanations make people wrongly think that they have enough information to modify the feature to obtain a desired outcome. But, since they have no knowledge on possible alternative values of the features, they actually cannot anticipate the outcome of the algorithm after modification of a given feature. Only transparent or counterfactual explanations significantly contributes to actionability. Even if the perceived actionability is equivalent between these two explanation modes, the transparent algorithm enables better performance of assessment tests related to actionability.

As above, Shapley explanations seem to overestimate perceived generalizability compared with generalizability test assessment scores. Surprisingly, counterfactual explanations may significantly decrease the ability of individuals to predict events compared to the situation where there is no explanation at all. Counterfactual explanations may distract users' attention by focusing on actionable variables to the extent that they significantly weaken people's ability to generalize. Counterfactual explanations can thus become tools of manipulation by withholding information and diverting users' attention. In terms of generalizability (both perceived and tested), transparent explanations perform much better than post-hoc explanations. Transparent explanations provide a global understanding on the mechanism behind the algorithm as opposed to post-hoc explanations that are too local to enable the generalization ability.

|  | None (baseline) | Transparent | Post-hoc Shapley | Post-hoc counterfactual | ANOVA |
|---|---|---|---|---|---|
| **Perceived understandability** | 4.06 | 4.9 pvalue < 0.01 | 4.45 pvalue = 0.07 | 3.98 pvalue = 0.7 | pvalue = 1e-5 Fvalue = 7.74 |
| **Tested understandability** | 1.06 | 1.36 pvalue < 0.01 | 1.52 pvalue < 0.01 | 1.16 pvalue = 0.3 | pvalue = 2e-5 Fvalue = 8.49 |
| **Perceived actionability** | 3.97 | 4.84 pvalue < 0.01 | 4.46 pvalue < 0.01 | 4.86 pvalue < 0.01 | pvalue = 2e-6 Fvalue = 10.2 |
| **Tested actionability** | 0.49 | 1.36 pvalue < 0.01 | 0.55 pvalue < 0.6 | 1.05 pvalue < 0.01 | pvalue = 3e-19 Fvalue = 33,3 |
| **Perceived generalizability** | 3.93 | 5.02 pvalue < 0.01 | 4.31 pvalue = 0.04 | 4.36 pvalue = 0.02 | pvalue = 1e-7 Fvalue = 12.3 |
| **Tested generalizability** | 0.78 | 1.22 pvalue < 0.01 | 0.8 pvalue = 0.8 | 0.56 pvalue = 0.02 | pvalue = 9e-11 Fvalue = 17.6 |

| Table 2. | Interpretability dimensions and trust averages within each scenario (t-test with H0: means equality with baseline scenario)[2] |
|---|---|

# Conclusion

Our results provide quantitative evidence of the limitations of post-hoc explanations of black-box-models compared to transparent ML models. While post-hoc explanations may look similar to transparent algorithm in terms of perceived dimensions of interpretability, we have seen that behavioral interpretability is strongly weakened by post-hoc explanations of a black-box model. Post-hoc explanations tend to give partial and biased information on the underlying mechanism of the algorithm, which tends to actually mislead the participants while they overestimate their capacity to interpret the decision in a declarative basis (Rudin 2019). Even though transparent model is not always possible in some use cases (e.g. in image detection), practitioners should prefer, as far as possible, inherently simple models or additive models rather than post-hoc explanations that may become tools of manipulation by withholding information and diverting users' attention. Our study shows that the opposition between self-reported indicators and tested behavioral indicators is key to understanding the pain points of post-hoc explanations of black-box models and providing a more comprehensive view of the dimensions of interpretability.

As the AI industry increasingly implements post-hoc explanations of black-box models, it is essential to produce critical results to question this trend. To do so, we have expanded the evaluation criteria by including the tested dimensions of interpretability. Interpretability as a solution to AI ethics thus has many limitations that it is fundamental to study in order to create a trustworthy AI systems.

# Current Work

Our first results are encouraging but their generalizability remains to be proven. Further study must be conducted to assess the generalizability of our results to different use cases (e.g. hiring algorithms) in different contexts. To do so, we are carrying out more work to answer to more questions:

- *Are our results robust to different situations?* To answer this question, we are repeating this study for 400 more people. For these additional 400 people, we simulate a discriminatory algorithmic decision that reject credit demand simply because demander is female. Since explanations of the AI decision can reveal the discriminatory nature of the decision, we measure the impact of the explanation on demanders through tested and perceived dimensions. The goal of the study is to assess the capacity of the explanations to detect discriminatory situations.
- *Are our results robust with respect to level of expertise of the users?* We need to complete the analysis with the socio-demographic variables that we collected in the preliminary phase of the interviews. For example, we can measure whether the limitations of the post-hoc explanations apply differently on people with low level of expertise (level of education, knowledge in AI).

# Acknowledgements

This research received funding from Good In Tech Chair, under the aegis of the Fondation du Risque in partnership with Institut Mines-Télécom and Sciences Po.

# Appendices

## *Appendix A: Introduction given to participants for each scenario*

*You apply for a loan from a bank. This bank uses an automatic artificial intelligence algorithm to decide whether to accept or refuse your application. The automatic algorithm uses the following 6 pieces of information about you to decide whether or not to grant you the loan:*
1. *Credit amount: €1500. This is the amount of the loan you are applying for.*
2. *Credit duration: 26 months. This is the time duration to pay the loan back.*

---

[2] Robustness check: Significance at 1% level is robust to outlier removal.

3. *Installment rate in percentage of disposable income: 4%. This is the proportion of your monthly income that will be spent to pay the loan back.*
4. *Number of years in employment: 5 years. This is the number of years you have been employed.*
5. *Property: Car. These are the assets you own.*
6. *Past loan approvals: 1. This is the number of credit application approvals you have already obtained with the same bank in the past. In your case, you have therefore obtained only one bank credit approval with this bank in the past.*

*Your application is processed by an artificial intelligence algorithm. This algorithm decides whether or not to grant you the loan based on these 6 variables and on past credit data from other bank customers. Based on the information you provided, the automatic algorithm decides not to grant you the loan. Your application for bank credit is therefore refused.*

### *Appendix B: Scenario 2 (Transparent algorithm)*

*To process your credit application, the automatic algorithm uses a points-based system. If the total number of points is greater than 210, then the algorithm accepts the credit application. Otherwise, the credit is refused. Below is a detailed explanation of how points are awarded:*

| | |
|---|---|
| 1. **Credit amount:**<br>• *Less than 1000 euros: 39 points*<br>• *Between 1000 and 2000 euros: 43 points*<br>• *Between 2000 and 3000: 47 points*<br>• *Between 3000 and 4000: 52 points*<br>• *Between 4000 and 5000: 37 points*<br>• *More than 5000: 42 points* | 4. **Number of years in employment:**<br>• *Without work: 37 points*<br>• *Less than 1 year: 31 points*<br>• *From 1 to 4 years: 37 points*<br>• *From 4 to 7 years: 42 points*<br>• *Over 7 years: 41 points* |
| 2. **Credit duration:**<br>• *Less than 6 months: 37 points*<br>• *Between 6 and 12 months: 19 points*<br>• *Between 12 and 18 months: 10 points*<br>• *Between 18 and 24 months: 7 points*<br>• *Between 24 and 30 months: 9 points*<br>• *Between 30 and 36 months: 5 points*<br>• *36 months and over: 0 point* | 5. **Property:**<br>• *Life insurance: 37 points*<br>• *Real estate: 48 points*<br>• *Car: 38 points*<br>• *No asset: 37 points* |
| 3. **Installment rate in percentage of disposal income:**<br>• *1%: 37 points*<br>• *2%: 36 points*<br>• *3%: 39 points*<br>• *4%: 36 points* | 6. **Number of credit application approvals in the past in the same bank:**<br>• *More than two credit approvals in the past with the same bank: 37 points*<br>• *At most one credit approval in the past with the same bank: 33 points* |
| **Table 3. Explanations given for transparent algorithm** | |

*Given your situation, you scored a total of 43 + 9 + 36 + 42 + 38 + 33 = 201 points.*
*Since your total number of points is less than 210 points, your credit application is refused.*

### *Appendix C: Scenario 3 (Shapley Post-hoc explanation)*

The graph below gives the variables that impacted the algorithmic decision. A green bar indicates that the variable had a positive impact on your credit application, i.e. the variable increases the chances of approval of your application. On the contrary, a red bar indicates that the variable had a negative impact on your credit application, i.e. the variable decreases the chances of approval of your application. Finally, the bigger the size of the bars, the greater the influence of the variable on the processing of your application.

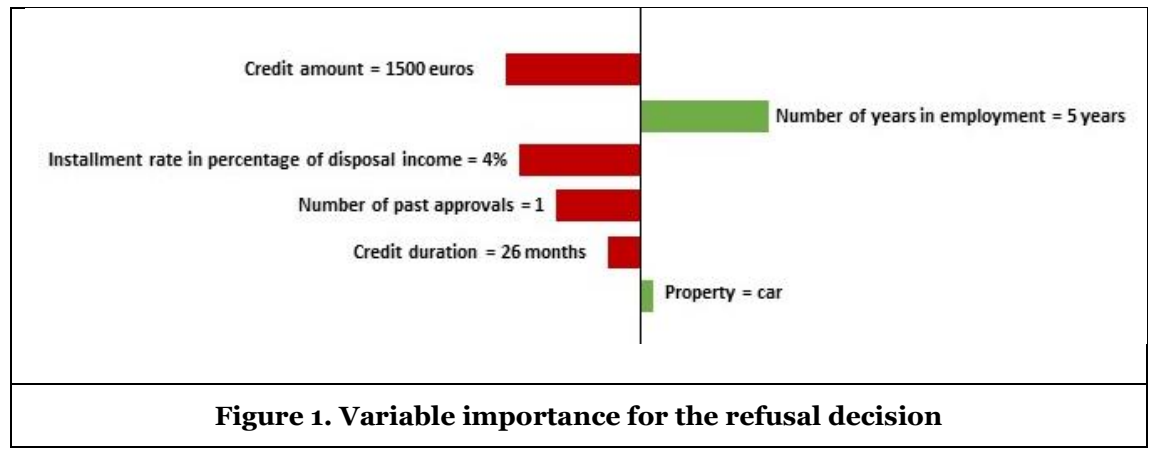

**Figure 1. Variable importance for the refusal decision**

*Appendix D: Scenario 4 (Counterfactual Post-hoc explanation)*

*- If the credit amount was between 3000 and 4000 euros, your credit application would have been accepted by the algorithm. As a reminder, your credit amount is currently 1500 euros.*
*- If the credit duration was reduced to less than 12 months, your credit request would have been accepted by the algorithm. As a reminder, your credit duration is currently 26 months.*

# References

"AI Explainability 360." (n.d.). (https://aix360.mybluemix.net/, accessed May 4, 2021).

Barredo Arrieta, A., Díaz-Rodríguez, N., del Ser, J., Bennetot, A., Tabik, S., Barbado, A., Garcia, S., Gil Lopez, S., Molina, D., Benjamins, R., Chatila, R., and Herrera, F. 2020. "Explainable Explainable Artificial Intelligence (XAI): Concepts, Taxonomies, Opportunities and Challenges toward Responsible AI," Information Fusion (58), Elsevier B.V., pp. 82–115. (https://doi.org/10.1016/j.inffus.2019.12.012).

Doshi-Velez, F., and Kim, B. 2017. Towards A Rigorous Science of Interpretable Machine Learning. (http://arxiv.org/abs/1702.08608).

Evermann, J., Rehse, J. R., and Fettke, P. 2017. "Predicting Process Behaviour Using Deep Learning," Decision Support Systems (100), Elsevier B.V., pp. 129–140. (https://doi.org/10.1016/j.dss.2017.04.003).

Gan, L., Wang, H., and Yang, Z. 2020. "Machine Learning Solutions to Challenges in Finance: An Application to the Pricing of Financial Products," Technological Forecasting and Social Change (153), Elsevier Inc. (https://doi.org/10.1016/j.techfore.2020.119928).

Hoffman, R. R., Mueller, S. T., Klein, G., and Litman, J. 2018. Metrics for Explainable AI: Challenges and Prospects. (http://arxiv.org/abs/1812.04608).

Jang, H. 2019. "A Decision Support Framework for Robust R&D Budget Allocation Using Machine Learning and Optimization," Decision Support Systems (121), Elsevier B.V., pp. 1–12. (https://doi.org/10.1016/j.dss.2019.03.010).

Kay, M., Matuszek, C., and Munson, S. A. 2015. "Unequal Representation and Gender Stereotypes in Image Search Results for Occupations," in Conference on Human Factors in Computing Systems - Proceedings (Vol. 2015-April), New York, NY, USA: Association for Computing Machinery, April 18, pp. 3819–3828. (https://doi.org/10.1145/2702123.2702520).

Kononenko, I. 2001. "Machine Learning for Medical Diagnosis: History, State of the Art and Perspective," Artificial Intelligence in Medicine (23:1), Elsevier, pp. 89–109. (https://doi.org/10.1016/S0933-3657(01)00077-X).

Krause, J., Perer, A., and Ng, K. 2016. "Interacting with Predictions: Visual Inspection of Black-Box Machine Learning Models," in Conference on Human Factors in Computing Systems - Proceedings, Association for Computing Machinery, May 7, pp. 5686–5697. (https://doi.org/10.1145/2858036.2858529).

Larson, J., Mattu, S., Kirchner, L., and Angwin, J. 2016. "How We Analyzed the COMPAS Recidivism Algorithm," ProPublica. (, accessed January 26, 2021).

Liem, C. C. S., Langer, M., Demetriou, A., Hiemstra, A. M. F., Sukma Wicaksana, A., Born, M. Ph., and König, C. J. 2018. Psychology Meets Machine Learning: Interdisciplinary Perspectives on Algorithmic Job Candidate Screening, Springer, Cham, pp. 197–253. (https://doi.org/10.1007/978-3-319-98131-4_9).

Lipton, Z. C. 2018. "The Mythos of Model Interpretability: In Machine Learning, the Concept of Interpretability Is Both Important and Slippery.," Queue (16:3), Association for Computing Machinery, pp. 31–57. (https://doi.org/10.1145/3236386.3241340).

Lu, J., Lee, D. (DK), Kim, T. W., and Danks, D. 2019. "Good Explanation for Algorithmic Transparency," SSRN Electronic Journal, Elsevier BV. (https://doi.org/10.2139/ssrn.3503603).

Lundberg, S. M., Allen, P. G., and Lee, S.-I. 2017. "A Unified Approach to Interpreting Model Predictions," in NIPS 2017. (https://github.com/slundberg/shap).

Mahmoudi, N., Docherty, P., and Moscato, P. 2018. "Deep Neural Networks Understand Investors Better," Decision Support Systems (112), Elsevier B.V., pp. 23–34. (https://doi.org/10.1016/j.dss.2018.06.002).

Martens, D. 2014. "Explaining Data-Driven Document Classifications," MIS Quaterly (38).

Martin, K. E. 2018. "Ethical Implications and Accountability of Algorithms," Journal of Business Ethics, Elsevier BV. (https://doi.org/10.2139/ssrn.3056716).

Miller, T. 2019. "Explanation in Artificial Intelligence: Insights from the Social Sciences," Artificial Intelligence, Elsevier B.V., pp. 1–38. (https://doi.org/10.1016/j.artint.2018.07.007).

Obermeyer, Z., Powers, B., Vogeli, C., and Mullainathan, S. 2019. "Dissecting Racial Bias in an Algorithm Used to Manage the Health of Populations," Science (366:6464), American Association for the Advancement of Science, pp. 447–453. (https://doi.org/10.1126/science.aax2342).

Poursabzi-Sangdeh, F., Goldstein, D. G., Hofman, J. M., Vaughan, J. W., and Wallach, H. 2018. "Manipulating and Measuring Model Interpretability," ArXiv, arXiv.

Rudin, C. 2019. "Stop Explaining Black Box Machine Learning Models for High Stakes Decisions and Use Interpretable Models Instead," Nature Machine Intelligence, Nature Research, pp. 206–215. (https://doi.org/10.1038/s42256-019-0048-x).

"SeldonIO/Alibi: Algorithms for Monitoring and Explaining Machine Learning Models." (n.d.). (https://github.com/SeldonIO/alibi, accessed May 4, 2021).